\documentclass[12pt,oneside]{kuthesisprop}

\usepackage{fancyhdr}
\usepackage{float}
\usepackage{fancybox}
\usepackage{graphicx} 
\pdfoutput=1 
\usepackage{amsmath}
\usepackage{url}
\usepackage{amsfonts}
\usepackage{amsthm}
\usepackage{setspace}
\usepackage{algorithm}
\usepackage{algorithmic}
\usepackage{array}
\usepackage{color}
\usepackage{courier}
\usepackage{verbatim}
\usepackage{cite}
\usepackage{subfigure}
\usepackage{booktabs}
\usepackage{bigstrut}
\usepackage{rotating}
\usepackage{multirow}
\usepackage{blindtext}
\usepackage{microtype}
\usepackage[english]{babel}
\usepackage{multirow}
\usepackage{mdwmath}
\usepackage{mdwtab}
\usepackage{caption}

\usepackage{pgfplots}
\pgfplotsset{compat=newest}
\pgfplotsset{plot coordinates/math parser=false}
\newlength\figureheight
\newlength\figurewidth

\newcommand{\m}[1]{\mbox{\boldmath $#1$}}

\newcommand{\head}[1]{\textnormal{\textbf{#1}}}

\newcommand{\as}[1]{\renewcommand{\arraystretch}{#1}}

\newcolumntype{L}[1]{>{\raggedright\let\newline\\\arraybackslash\hspace{0pt}}m{#1}}
\newcolumntype{C}[1]{>{\centering\let\newline\\\arraybackslash\hspace{0pt}}m{#1}}
\newcolumntype{R}[1]{>{\raggedleft\let\newline\\\arraybackslash\hspace{0pt}}m{#1}}

\begin{document}
\setcounter{secnumdepth}{4} 
\setcounter{tocdepth}{4} 

\prelimpages

\Title{Enhancement of Throat Microphone Recordings Using Gaussian Mixture Model Probabilistic Estimator}

\Author{Mehmet Ali Tu\u{g}tekin Turan}

\Year{August, 2013}

\Signature{ Assoc.~Prof.~Engin~Erzin }

\Signature{ Prof.~Murat~Tekalp }

\Signature{ Prof.~Levent~Arslan }

\titlepage

\thesissignaturepage

\dedication{ \vspace*{5cm}

\begin{center}
{\em\large In human life, you will find players of religion until the knowledge and proficiency in religion will be cleansed from all superstitions, and will be purified and perfected by the enlightenment of real science.}
\end{center}
\begin{center}
{\em {\Large \textbf{Mustafa Kemal Atat\"urk}}}
\end{center}

}


\abstract{

The throat microphone is a body-attached transducer that is worn against the neck. It captures the signals that are transmitted through the vocal folds, along with the buzz tone of the larynx. Due to its skin contact, it is more robust to the environmental noise compared to the acoustic microphone that picks up the vibrations through air pressure, and hence the all interventions. The throat speech is partly intelligible, but gives unnatural and croaky sound. This thesis tries to recover missing frequency bands of the throat speech and investigates envelope and excitation mapping problem with joint analysis of throat- and acoustic-microphone recordings. A new phone-dependent GMM-based spectral envelope mapping scheme, which performs the minimum mean square error (MMSE) estimation of the acoustic-microphone spectral envelope, has been proposed. In the source-filter decomposition framework, we observed that the spectral envelope difference of the excitation signals of throat- and acoustic-microphone recordings is an important source of the degradation in the throat-microphone voice quality. Thus, we also model spectral envelope difference of the excitation signals as a spectral tilt vector, and propose a new phone-dependent GMM-based spectral tilt mapping scheme to enhance throat excitation signal. Experimental evaluations are performed to compare the proposed mapping scheme using both objective and subjective evaluations. Objective evaluations are performed with the log-spectral distortion (LSD) and the wide-band perceptual evaluation of speech quality (PESQ) metrics. Subjective evaluations are performed with A/B pair comparison listening test. Both objective and subjective evaluations yield that the proposed phone-dependent mapping consistently improves performances over the state-of-the-art GMM estimators.

}

\oz{   

G\i rtlak mikrofonu, ses tellerindeki titre\c{s}imi g\i rtlaktan gelen sinyallerle beraber ileten ve kullanan ki\c{s}inin boynuna takt\i \u{g}\i~insan bedeniyle temas eden bir mikrofon t\"{u}r\"{u}d\"{u}r. Bu ba\u{g}lant\i~sayesinde, titre\c{s}imleri havadan alan akustik mikrofonlara nazaran g\"{u}r\"{u}lt\"{u} gibi \c{c}evresel etmenlere kar\c{s}\i~daha g\"{u}rb\"{u}z bir ileti\c{s}im sa\u{g}lar. G\i rtlak mikrofonu ile kaydedilen sesler k\i smen de olsa anla\c{s}\i lmas\i na ra\u{g}men, do\u{g}al olmayan ve kula\u{g}\i~rahats\i z edici bir yap\i dad\i r. \.{I}\c{s}te bu \c{c}al\i \c{s}ma g\i rtlak mikrofonlar\i ndaki \"{u}retilemeyen frekans aral\i klar\i n\i~geri kazanabilmeyi ama\c{c}larken ayn\i~zamanda sesin kaynak ve s\"{u}zge\c{c} k\i s\i mlar\i n\i~do\u{g}ru tahmin edebilme sorununu, g\i rtlak ve akustik kay\i tlar\i~m\"{u}\c{s}terek bir \c{s}ekilde \c{c}\"{o}z\"{u}mleyerek irdelemektedir. Bu ba\u{g}lamda, ortalama kare hatas\i n\i~en aza indirerek, ses birimlerine ba\u{g}l\i~Gauss kar\i \c{s}\i m modeli tabanl\i~bir kestirici sistemi \"{o}ne s\"{u}r\"{u}lm\"{u}\c{s}t\"{u}r. Kaynak--s\"{u}zge\c{c} ayr\i \c{s}t\i rmas\i~\c{c}er\c{c}evesinde, g\i rtlak ve akustik s\"{u}zgecinin g\"{o}r\"{u}ngesel farkl\i l\i klar\i n\i n, g\i rtlak mikrofonundan gelen ses kalitesini d\"{u}\c{s}\"{u}ren \"{o}nemli bir etmen oldu\u{g}unu g\"{o}zlemledik. Bu sebepten \"{o}t\"{u}r\"{u}, yukar\i da bahsedilen fark\i~g\"{o}r\"{u}ngesel e\u{g}im vekt\"{o}r\"{u} olarak modelleyip, g\i rtlak s\"{u}zgecini iyile\c{s}tirici bir sistemi ayr\i ca \"{o}ne s\"{u}rd\"{u}k. Ortaya konulan sistemlerin katk\i lar\i n\i~yorumlayabilmek i\c{c}in hem nesnel hem de \"{o}znel deneyler tasarlad\i k. Nesnel deneyler, logaritmik g\"{o}r\"{u}nge tahribat\i~ve ses kalitesinin alg\i sal de\u{g}erlendirilmesi k\i staslar\i~\"{u}zerinden incelendiler. Bununla birlikte, \"{o}znel de\u{g}erlendirmeler ise A/B e\c{s} kar\c{s}\i la\c{s}t\i rma deneyi \c{s}eklinde tatbik edildi. Hem nesnel hem de \"{o}znel deneyler g\"{o}sterdi ki \"{o}ne s\"{u}r\"{u}len ses birimi tabanl\i~kestirimler, halihaz\i rda bulunan Gauss kar\i \c{s}\i m modeli tabanl\i~kestirimlere g\"{o}re tutarl\i~bir \c{s}ekilde iyile\c{s}tirmeler sa\u{g}lamaktad\i r.

}

%
%

\tableofcontents

\listoftables

\listoffigures

\abbreviations{
\begin{tabular}{lll}
$ TM $ & : & Throat-Microphone \\
$ AM $ & : & Acoustic-Microphone \\
$ GMM $ & : & Gaussian Mixture Model \\
$ MMSE $ & : & Minimum Mean Square Error \\
$ LSD $ & : & Log-Spectral Distortion \\
$ PESQ $ & : & Perceptual Evaluation of Speech Quality \\
$ NAM $ & : & Non-Audible Murmur \\
$ LSF $ & : & Line Spectrum Frequency \\
$ LPC $ & : & Linear Predictive Coding \\
$ HMM $ & : & Hidden Markov Model \\
$ VQ $ & : & Vector Quantization \\
$ EM $ & : & Expectation Maximization \\
$ SM $ & : & Soft Mapping \\
$ HM $ & : & Hard Mapping \\
$ POF $ & : & Probabilistic Optimum Filter \\
$ PDSM $ & : & Phone-Dependent Soft Mapping \\
$ PDHM-G $ & : & Phone-Dependent Hard Mapping with the GMM Classifier \\
$ PDHM-M $ & : & Phone-Dependent Hard Mapping with the HMM Phone Recognition \\
$ PDHM-T $ & : & Phone-Dependent Hard Mapping with the True Phone Class \\
\end{tabular}
}

\textpages
\chapter{Introduction}
\label{ch:ch1_intro}

\section{Non-Acoustic Sensors}
\label{subsec:throatMic}

Multi-sensor configurations have recently been applied to speech enhancement problem which mainly aims to obtain high performance quality of speech. Environmental effects such as background noise or wind turbulence motivated researchers to use different mediums such that speech can be spread other than by means of air. Other mediums, such as human tissue, infrared ray, light wave, and laser also can be used to detect the non-air conducted speech or acoustical vibrations. A few type of non-acoustic sensors, i.e. sensors that don't not use the air, are developed due to this reason, however, their application are limited since detection materials are usually difficult to obtain ~\cite{Wang2009}. 

The traditional acoustic microphones use air as a medium of sound conduction. Thus, it is ineffective in extreme conditions such as acoustic noise. On the other hand, piezoelectric transducers in non-acoustic sensors can pick up voice signals by absorbing the vibrations generated from human body. Thus, all of the non-acoustic sensors are insensitive to environmental conditions. However, they only capture low-frequency portion of sound and may distort speech signals due to their low energy static noise. This causes a reduced frequency bandwidth unfortunately. Another advantage of them is the ability to reveal certain speech attributes lost in the noisy acoustic signal; for example, low-energy consonant voice bars and nasality excitation. These sensors provide measurements of functions of the glottal excitation and, more generally, of the vocal tract articulator movements that are relatively resistant to acoustic disturbances \cite{Quatieri2006}.

\begin{figure}[!ht]
\begin{center}
\includegraphics[scale=1]{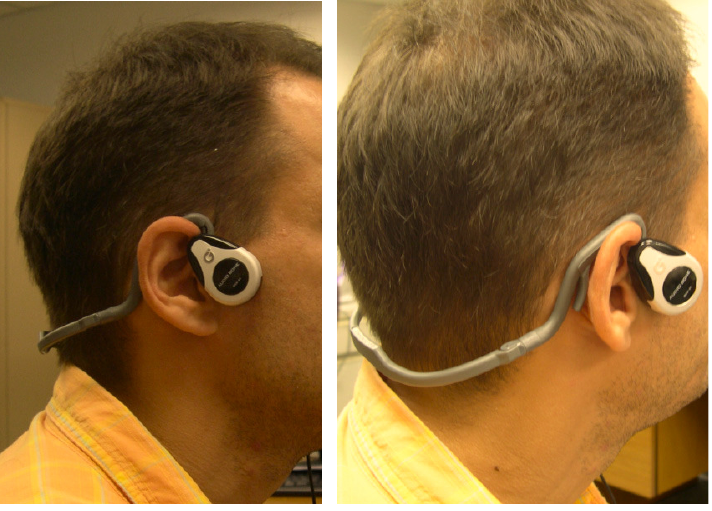}
\caption{Bone-Conducting Headset \cite{Lindeman2007}}.
\label{fig:bonecond}
\end{center}
\end{figure}

One of the early studies about non-acoustic sensors have been originated from human auditory system. Since hearing is actualized both air and bone conduction pathways, a playback from recorded speech is perceived different to us. Likewise, bone conducting microphones, see Figure~\ref{fig:bonecond}, can catch signals from the inner ear through the bones of the skull \cite{Lindeman2007}. It has many advantages over air conduction owing to its robustness to noise. For example, it can be used for getting information whether the user is talking or not \cite{Cutler2000}. It is commonly used in military environments such as headphone-based communication interfaces. Bone conduction sensors provides an effective transmission without interfering with hearing protection devices. In quiet environments, the soldier could receive radio communications through bone conduction without obscuring the ears, thereby maintaining full awareness of the surrounding acoustic environment \cite{MacDonald2006}. However, one of the main disadvantages of current bone conduction systems is that they are restricted to single channel operation. Due to this reason, it is typically used for enhancements as a supplementary speech source. In \cite{Zheng2003}, combining the two channels from the air- and bone- conductive microphone, it is possible to remove background speech.

\begin{figure}[!ht]
\begin{center}
\includegraphics[scale=1.25]{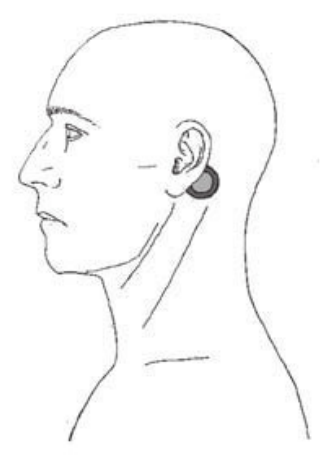}
\caption{Non-Audible Murmur (NAM) Microphone \cite{Toda2012NAM}}
\label{fig:nam}
\end{center}
\end{figure}

Another favored non-acoustic sensor is Non-Audible Murmur (NAM) microphone that is attached behind the speaker's ear (see Figure~\ref{fig:nam}). The specialty of NAM is the ability of capturing very quietly uttered speech that cannot be heard by listeners through human tissue \cite{Heracleous2005}. Since it captures inaudible speech produced without vibration of glottis, it is difficult to identify the differences between whisper and NAM speech. The principle behind NAM sensor is based on medical stethoscope used for monitoring internal sounds of human body. Similarly the NAM microphone is mainly used for privacy purposes while communicating with speech recognition engines. The NAM users don't pay so much effort owing to quiet utterance so this provides communication without hearing from others. On the other hand, it can be useful for diseased people who have physical difficulties in speech \cite{Heracleous2003}.

\begin{figure}[!ht]
\centering
\includegraphics[scale=0.55]{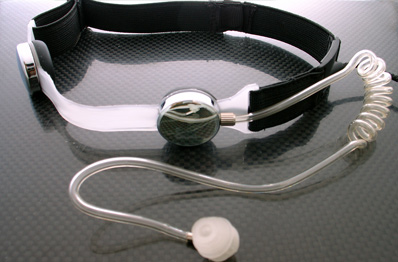}
\caption{Throat Microphone (TM) \cite{IASUS}}
\label{fig:IASUS}
\end{figure}

Another specialized non-acoustic sensor is throat microphone (TM) that have been used in military applications and radio communication for several years (see Figure~\ref{fig:IASUS}). It can capture speech signals in the form of vibrations and resonances of vocal cords through skin-attached piezoelectric sensors \cite{Shahina2007}. Since the signals are acquired from the throat, they have lower bandwidth speech signals compared to acoustic-microphone (AM) recordings. Like other non-acoustic sensors, the TM recordings are significantly more robust to environmental noise conditions, however they suffer from the perceived speech quality \cite{Erzin2009}. Since TM recordings are strongly robust and highly correlated with the acoustic speech signal, they are attractive candidates for robust speech recognition applications under adverse noise conditions, such as airplane, motorcycle, military field, factory or street crowd environments. Likewise with the NAM, it can be used for patients who have lost their voices due to injury or illness, or patients who have temporary speech loss after a tracheotomy. One of the biggest problem is the quality degradation of TM speech caused by deficiency of oral cavity transmission such as lack of lip radiation. This problem can be handled with synchronous analysis of AM and TM speech. Moreover, the TM is useful in terms of its distinct formant-like structures that serve as acoustic cues that can be used to resolve the highly confusable voiced phones into classes based on the place of articulation \cite{Shahina2007}. Since the TM conveys more information about the AM speech characteristics among other non-acoustic sensors, we aim to improve its perceived quality by mapping its spectral features in this thesis.

\section{Scope}

It is emphasized in previous section that the TM is lack of perceptual quality because of its muffled speech. In the literature, there are a few attempts which aim to enhance the quality of throat-only speech. The published works mostly deal with robust speech recognition by means of the TM in the extreme environments. In this research, we target to enhance the naturalness and the intelligibility of the TM speech by mapping not only its filter but also its excitation spectra closer to the one that belongs to the AM speech via Gaussian mixture model (GMM) probabilistic estimators. This mapping is trained using simultaneously recorded acoustic- and throat-microphone speech and is formulated by context independent and dependent methods. Moreover, the outcome of this work may expand to a linear system that uses the direct filtering so that it can be speaker independent. Thus, all of the proposed schemes try to improve the understandability of throat-only speech.

\section{Related Work}

In one of the early studies, Viswanathan et al.~presented a two sensor system involving an accelerometer mounted on the speaker's throat and a noise-canceling microphone is located close to the lips \cite{Roucos1986}. Close talking first- and second-order differential microphones are designed to be placed close to the lips where the sound field has a large spatial gradient and the frequency response of the microphone is flat. Second-order differential microphones using a  single element piezoelectric transducer have been suggested for use in very noisy environment of aircraft communication systems to enhance a noisy signal for improved speech recognition. 

A device that combines a close-talk and a bone-conductive microphone is proposed by the Microsoft research group for speech detection using a moving-window histogram \cite{Zheng2003}. They tried to handle non-stationary noises in both automatic speech recognition and audio enhancement. Note that the bone sensor is highly insensitive to ambient noise and provides robust speech activity detection. They showed that such devices can be used to determine whether the speaker is talking or not, and furthermore, the two channels can be combined to remove overlapping noise. It works by training a piecewise linear mapping from the bone signal to the close-talk signal. One drawback of this approach is that it requires training for each speaker. This problem can be solved with a technique called \textit{Direct Filtering} that does not require any training. It is based on learning mappings in a maximum likelihood framework and investigated in \cite{Liu2004}. Later, direct filtering is improved to deal with the environmental noise leakage into the bone sensor and with the teeth-clack problem \cite{Liu2005}. 

The use of non-acoustic sensors in multi-sensory speech processing has been studied for speech enhancement, robust speech modeling and improved speech recognition \cite{Subramanya2006,Quatieri2006,Jou2005,Erzin2009}. Multi-sensory speech processing for noisy speech enhancement and improved noise robust speech recognition are discussed in \cite{Subramanya2006,Subramanya2005,Subramanya2005a}. In these works, Subramanya et al.~proposed an algorithm based on the SPLICE technique and a speech detector based on the energy in the bone channel. In another multi-sensory study, speech recorded from throat and acoustic channels is processed by parallel speech recognition systems and later a decision fusion yields robust speech recognition to background noise \cite{Dupont2004}. Due to the approximation of the sensor to the voice source, the signal is naturally less exposed to background noise. In \cite{Dupont2004}, Dupont et al.~proposed to use the information from both signals by combining the probability vectors provided by both models.

Graciarena et al.~proposed estimation of clean acoustic speech features using the probabilistic optimum filter (POF) mapping with combined throat and acoustic microphone recordings \cite{Graciarena2003}. The POF mapping is a piecewise linear transformation applied to noisy feature space to estimate the clean ones \cite{Neumeyer1994}. It maps the temporal sequence of noisy mel-cepstral features from the standard and the throat microphone. Thus, this mapping allows for an optimal combination of the noisy and the throat speech. In \cite{Erzin2009}, Erzin developed a framework to define a temporal correlation model between simultaneously recorded throat- and acoustic-microphone speech. This framework aims to learn joint sub-phone patterns of throat and acoustic microphone recordings that define temporally correlated neighborhoods through a parallel branch hidden Markov model (HMM) structure. The resulting temporal correlation model is employed to estimate acoustic features, which are spectrally richer than throat features, from throat features through linear prediction analysis. The throat and the estimated acoustic microphone features are then used in a multi-modal speech recognition system.

Non-acoustic sensors can reveal speech attributes that are lost in the noisy acoustic signal such as, low-energy consonant voice bars, nasality, and glottal excitation. Quatieri et al.~investigate methods of fusing non-acoustic low-frequency and pitch content with acoustic-microphone content for low-rate coding of speech \cite{Quatieri2006}. By fusing non-acoustic low-frequency and pitch content with acoustic-microphone content, they achieved significant intelligibility performance gains using the diagnostic rhyme test across a variety of environments.

Although throat-microphone recordings are robust to acoustic noise and reveal certain speech attributes, they often lack naturalness and intelligibility. There have been a few attempts in the literature that improve the perceived speech quality of non-acoustic sensor recordings. A neural network based mapping of the speech spectra from throat-microphone to acoustic-microphone recordings has been investigated in \cite{Shahina2007}. This neural network is used to capture the speaker-dependent functional relationship between the feature vectors, i.e.~cepstral coefficients, of the speech signals. Moreover, speech spectra mapping techniques have been also studied extensively for the artificial bandwidth extension of telephone speech \cite{Jax2003,Yagli2012}. This method aims to estimate wide-band speech (50 Hz - 7 kHz) from narrow-band signals (300 Hz - 3.4 kHz). Applying the source-filter model of speech, many existing algorithms estimate vocal tract filter parameters independently of the source signal. In another study \cite{Kondo2006}, the transfer characteristics of bone-conducted and acoustic-microphone speech signals are modeled as dependent sources, and an equalizer, which is trained using simultaneously recorded acoustic and bone-conducted microphone speech, has been investigated to enhance bone-conducted speech. 

Speech enhancement of non-acoustic sensor recordings also employs techniques used for voice conversion \cite{Stylianou1998,Toda2007} and artificial bandwidth extension \cite{Jax2003,Yagli2012} to improve naturalness and intelligibility of the speech signal. One widely used framework for enhancement of the non-acoustic sensor recordings is the source-filter decomposition, which breaks down the problem into two, namely the enhancement of the excitation (source) and the spectral envelope (filter).

Enhancement of the spectral envelope has been both studied for the speech conversion and the artificial bandwidth extension problems. Stylianou et al.~\cite{Stylianou1998} presented one of the early works on continuous probabilistic mapping of the spectral envelope for the voice conversion problem which is simply defined as modifying the speech signal of one speaker (source
speaker) so that it sounds like be pronounced by a different speaker (target speaker). Their  contribution includes the design of a new methodology for representing the relationship between two sets of spectral envelopes. Their proposed method is based on the use of a Gaussian mixture model of the source speaker spectral envelopes. The conversion itself is represented by a continuous parametric function which takes into account the probabilistic classification provided by the mixture model. Later Toda et al.~\cite{Toda2007} improved the continuous probabilistic mapping by incorporating not only static but also dynamic feature statistics for the estimation of a spectral parameter trajectory. Furthermore, they tried to deal with the over-smoothing effect by considering a global variance feature of the converted spectra.

Enhancement of the excitation has been studied on domain specific problems and not as widely as the enhancement of spectral envelope. Recently, conversion methodologies from NAM to acoustic and whispered speech have been developed to improve voice quality and intelligibility of NAM speech \cite{Toda2012}. In \cite{Toda2012}, spectral and excitation features of acoustic speech are estimated from the spectral feature of NAM. Since NAM lacks fundamental frequency information, a mixed excitation signal is estimated based on the estimated fundamental frequency and aperiodicity information from NAM. The converted speech reported to suffer from unnatural prosody because of the difficulty of estimating the fundamental frequency of normal speech. In another study \cite{Kondo2006}, the transfer characteristics of bone-conducted and acoustic-microphone speech signals are modeled as dependent sources, and an equalizer, which is trained using simultaneously recorded acoustic and bone-conducted microphone speech, has been investigated to enhance bone-conducted speech. Since the transfer function of the bone-conduction path is speaker and microphone dependent, the transfer function should be individualized for effective equalization. Then, Kondo et al.~\cite{Kondo2006} propose a speaker-dependent short-term FFT based equalization with extensive smoothing. In the bandwidth extension framework, the extension of the excitation signal has been performed by modulation, which attains spectral continuation and a matching harmonic structure of the baseband \cite{Jax2003}. In other words, this method guarantees that the harmonics in the extended frequency band always match the harmonic structure of the baseband. Moreover, their pitch-adaptive modulation reacts quite sensitive to small errors of the estimate of the pitch frequency.

\section{Contributions}

In this thesis, we have the following contributions over state-of-art techniques that are investigated in \cite{Shahina2007,Jax2003,Yagli2012,Kondo2006}: 

\begin{itemize}

\item
The main contribution of this work is the context-dependency of estimations, which is set at the phone level. We observe significant improvements when the true phone-context is available for the both envelope and excitation mappings. Based on this observation, we investigate some phone-dependent mapping schemes in the presence of predicted phone-context. 

\item
We introduce a new phone-dependent GMM-based spectral envelope mapping scheme to enhance TM speech using joint analysis of TM and AM recordings. The proposed spectral mapping scheme performs the minimum mean square error (MMSE) estimation of the AM spectral envelope within the phone class neighborhoods. Objective and subjective experimental evaluations indicate that the phone-dependent spectral mapping yields perceivable improvements over the state-of-the-art context independent schemes.

\item 
We also observed that the spectral envelope difference of the excitation signals of TM and AM recordings is an important source of the degradation for the TM voice quality. Thus, we model spectral envelope difference of the excitation signals as a spectral tilt vector, and propose a new context-dependent probabilistic spectral tilt mapping scheme based on MMSE estimation. We consider that incorporating temporal dynamics of the spectral tilt to the probabilistic mapping expectedly attain further improvements for TM speech enhancement. 

\end{itemize}

\section{Organization}
The organization of this thesis as follows: In chapter~\ref{ch:ch2_enhancement} we introduce the proposed throat-microphone speech enhancement system. By doing this, we first review the well-known source-filter model of human speech system and try to associate the proposed mappings to the statement of this theory. Chapter~\ref{ch:ch3_experiments} discusses the experimental evaluations using both objective and subjective results. Finally, chapter~\ref{ch:conc} includes the conclusion and future direction of this study.
\chapter{Enhancement System}
\label{ch:ch2_enhancement}

\section{System Overview}
\label{subsec:overview}

We start to learning part by source-filter separation of TM and AM speech. We use line spectral frequency (LSF) features to represent envelope spectra (filter) of the speech signal as training features, the prediction coefficients are firstly converted to the LSFs. We use Gaussian mixtures as probabilistic estimator model that is a parametric probability density function represented as a weighted sum of Gaussian component densities. The discourse about this model and our proposed modifications on it are discussed in Section~\ref{subsec:GMM}.

For the envelope mapping, joint distribution of AM and TM spectral envelopes are modeled and we define tilt features $D(b)$ based on the spectral envelope difference of the TM and AM excitation signals for the enhancement of source (excitation). We also use cepstral feature vectors $C_T(n)$ to constitute the observable part of the excitation mapping. At the end, GMM-based training is applied to the time-aligned TM and AM features. Details about the calculations of these features are comprehensively examined in Section~\ref{subsec:EnhFrame} -- \ref{subsec:ExcMapping}. 

\begin{figure}[H]

\hspace{-6em}
\includegraphics[scale=0.37]{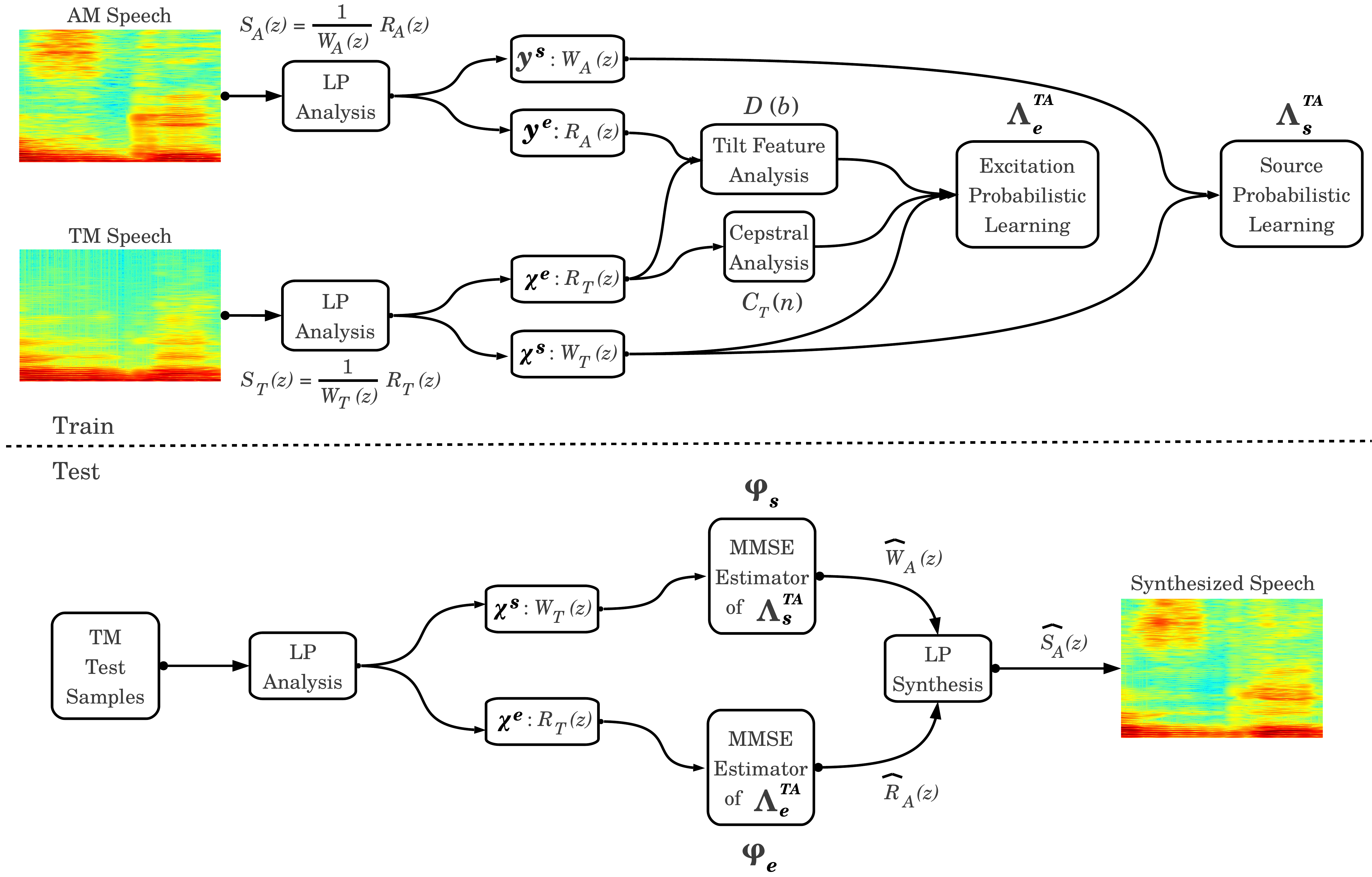}
\caption{Block Diagram of Enhancement System}
\label{fig:architecture}

\end{figure}

In the test stage, the throat-microphone test recordings are separated into source $R_T(z)$ and filter $W_T(z)$ through linear prediction analysis. The estimated acoustic filter $\hat{W}_A(z)$ is extracted from the throat filter $W_T(z)$ and the estimated acoustic source $\hat{R}_A(z)$ is computed using the throat cepstral feature vectors $C_T(n)$ and the throat filter $W_T(z)$ via different mapping schemes based on the minimum mean square error (MMSE) approach. Then, the enhanced throat-microphone recordings are synthesized using these estimated source and filter. The summary of whole system is depicted in Figure~\ref{fig:architecture}.

\section{Source-Filter Separation}
\label{subsec:sourceFilter}

\begin{figure}[!ht]
\centering
\includegraphics[scale=0.5]{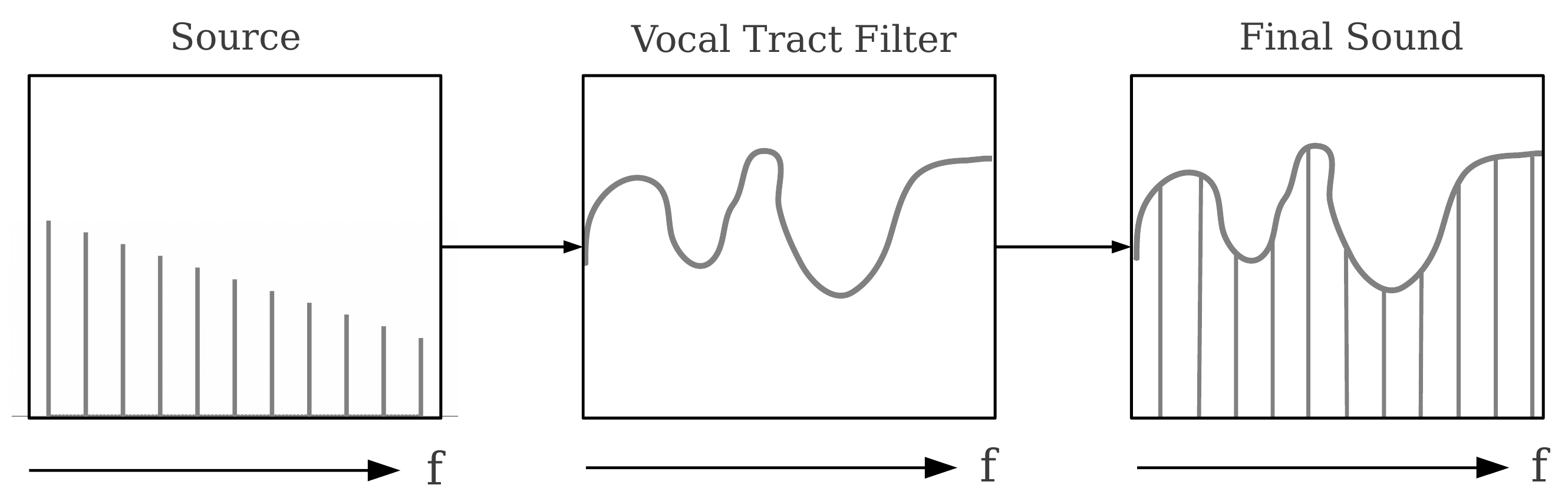}
\caption{Source-Filter Model of Speech}
\label{fig:sourcefilter}
\end{figure}

In 1960, Gunnar Fant, from Royal Institute of Technology (KTH), observed that the glottis and the vocal tract are totally distinct and both of them can be designed independently of each other. The \textit{source-filter model} has been introduced based on the fact that speech is produced by an excitation signal generated in our throat, which is modified by resonances produced by different shapes of our vocal, nasal and pharyngeal tracts \cite{Dutoit2009}. In other words, the speech signal can be decomposed as a source passed through a linear time-varying filter where the source represents the air flow at the vocal cords, and the filter represents the resonances of the vocal tract that changes over time (see Figure~\ref{fig:sourcefilter}).

Speech sounds are generally divided into two main groups, namely voiced and unvoiced. Voiced sounds such as vowels, liquids, glides and nasals are produced when the vocal tract is excited by air pressure due to opening and closing of the vocal cords. On the other hand, unvoiced sounds are produced by creating turbulent air flow, which acts as a random noise excitation of the vocal tract. According to source-filter model of speech, the excitation of a voiced sound is a quasi-periodic sequence of discrete glottal pulses whose fundamental frequency determines the perceived pitch of the voice whereas its unvoiced counterpart behaves like discrete-time noise signal with flat spectrum.

The question about the estimation of speech parameters are solved by means of linear predictive (LP) analysis. It simply states that the current features can be predicted using weighted sum of the past ones. By doing this, it analyses the speech by estimating its formants, resonances of the vocal tract, and then eliminates its effect from the speech signal. The process of removing the formants is called inverse filtering, and the remaining signal is called the residue \cite{Furui1989}. Then, LP analysis realizes the reverse of this process to create a source signal, and uses the formants to construct a filter and run the source through this filter which results in actual speech. More clearly, this linear system is described by an all-pole filter of the form \cite{Huang2001}: 
\begin{equation}
H(z) = \frac{S(z)}{E(z)} = \frac{1}{1 - \sum\limits_{k = 1}^{p}a_{k}z^{-k}}
\end{equation}
In this linear system, the speech samples $ s[n] $ are related to the excitation $ e[n] $ by the difference equation 
\begin{equation}
\label{eqn:difeq}
s[n] = \sum\limits_{k = 1}^{p}a_{k}s[n-k] + e[n].
\end{equation}
A linear predictor with prediction coefficients, $ \alpha_{k} $, is defined as a system whose output is
\begin{equation}
\tilde{s}[n] = \sum\limits_{k = 1}^{p}\alpha_{k}s[n-k]\:,
\end{equation}
and the prediction error, defined as the amount by which $ \tilde{s}[n] $ fails to exactly predict sample $ s[n] $ is
\begin{equation}
d[n] = s[n] - \tilde{s}[n] = s[n] - \sum\limits_{k = 1}^{p}\alpha_{k}s[n-k].
\end{equation}
It follows that the prediction error sequence is the output of an FIR linear system whose system function is
\begin{equation}
\label{eqn:inverseFilter}
A(z) = 1 - \sum\limits_{k = 1}^{p}\alpha_{k}z^{-k} = \frac{D(z)}{S(z)}
\end{equation}
It can be clearly seen that the speech signal obeys the source-filter model, and if $ \alpha_{k} = a_{k} $ and $ d[n] = e[n] $. Thus, the prediction error filter $ A(z) $ becomes an inverse filter for the system $ H(z) $ that is
\begin{equation}
H(z) = \frac{1}{A(z)}
\end{equation}

After all these equations, a simple question arises about the determination of the predictor coefficients $ \{\alpha_{k}\} $ from the speech \cite{Rabiner2007}. The ubiquitous solution is to find a set that minimizes the mean square error of the speech waveform. There are some important motivations behind this approach, such as the predictor coefficients that comes from mean-squared minimization are identical to the coefficients of difference equation in \eqref{eqn:difeq}. Moreover, if $ \alpha_{k} $ is equal to $ a_{k} $ and $ d[n] $ is equal to $ e[n] $, $ d[n] $ should be almost equal to impulse train except at isolated samples spaced by the current pitch period \cite{Rabiner2007}. Eventually, one approach to computing the prediction coefficients is based on the \textit{covariance method} that is direct solution of well-known Yule-Walker equations. However, the most widely used method is called \textit{autocorrelation method} because the covariance function, i.e. Yule-Walker equations, has no specific range \cite{Huang2001}. In other words, we can use a analysis window in the autocorrelation method which provides zero prediction error outside the window interval \cite{Furui1989}. Briefly, the latter method is based on the short-time autocorrelation function which is the inverse discrete Fourier transform of the magnitude-squared of the short-time Fourier transform of the windowed speech signal \cite{Rabiner2007}.

Since the important parameter of this linear model is the prediction coefficients $ \{\alpha_{k}\} $, there is a considerable amount of equivalent representations for these coefficients. All of them have a distinct characteristic that is very important especially in speech coding because of the parameter quantization \cite{Dutoit2009}. The first one comes from the roots of inverse filter in \eqref{eqn:inverseFilter}. It is clear that this representation is a polynomial, the coefficients can be interpreted as zeros of $ A(z) $ that are poles of $ H(z) $ by definition. Due to the this fact, the filter of LP model is also called as all-pole filter. However, the roots are quite sensitive to the quantization errors which makes the system unstable. Therefore, a much more robust option is proposed by Itakura in \cite{Itakura1975} called Line Spectrum Frequencies (LSF). 

LSFs collectively describe the two resonance conditions arising from an interconnected tube model of the human vocal tract. This includes mouth shape and nasal cavity, and forms the basis of the underlying physical relevance of the linear prediction representation \cite{Vaseghi2006}. The two resonance conditions describe the modeled vocal tract as being either fully open or closed at the glottis respectively. The resonances of each condition give rise to odd and even line spectral frequencies respectively, and are provided into a set of LSFs which have monotonically increasing value \cite{McLoughlin2009}. In reality, however, the human glottis opens and closes rapidly during speech so it is neither fully closed nor open. Hence, actual resonances occur at frequencies located somewhere between the two extremes of each LSF condition. Nevertheless, this relationship between resonance and LSF position leads a significant physical basis to the representation.

In more detail, LSFs are a direct mathematical transformation of the set of LP parameters, and are used within many speech compression systems. It is very popular due to their excellent quantization characteristics and consequent efficiency of representation. To explain it in more detail, we can define two ($p$ + 1)-th order polynomials related to $ A(z) $, named $ P(z) $ and $ Q(z) $. These are referred to as antisymmetric (or inverse symmetric) and symmetric parts based on observation of their coefficients. The polynomials represent the interconnected tube model of the human vocal tract and correspond respectively to complete closure and opening at the source part of the interconnected tubes. In the original model, the source part is the glottis, and is neither fully open nor closed during the period of analysis, and thus the actual resonance conditions encoded in $ A(z) $ are a linear combination of the two boundaries \cite{McLoughlin2009}. In fact this is simply stated as
\begin{equation}
A(z) = \frac{P(z) + Q(z)}{2}\:,
\end{equation}
where $ A(z) $ is the LP polynomial that is derived in \eqref{eqn:inverseFilter}.

The two polynomials are created from the LPC polynomial with an extra feedback term being positive to model energy reflection at a completely closed glottis, and negative to model energy reflection at a completely open glottis
\begin{eqnarray}
P(z) = A(z) - z^{-(p+1)}A(z^{-1}) \\
Q(z) = A(z) + z^{-(p+1)}A(z^{-1})
\end{eqnarray}

The roots of these two polynomials are the set of line spectral frequencies, $ \omega_k $ that can be located on the unit circle in the z-plane if the original LPC filter was stable and alternate around the unit circle. Remember that any equivalent size set of roots that alternate in this way around and on the unit circle will represent a stable LPC filter. In practice, LSF are useful because of sensitivity (a quantization of one coefficient generally results in a spectral change only around that frequency) and efficiency (LSF results in low spectral distortion). At the end, as long as the LSFs are ordered, the resulting LPC filter is stable.

\section{Gaussian Mixture Model (GMM)}
\label{subsec:GMMintro}

\begin{figure}[H]
\begin{center}
\includegraphics[scale=0.5]{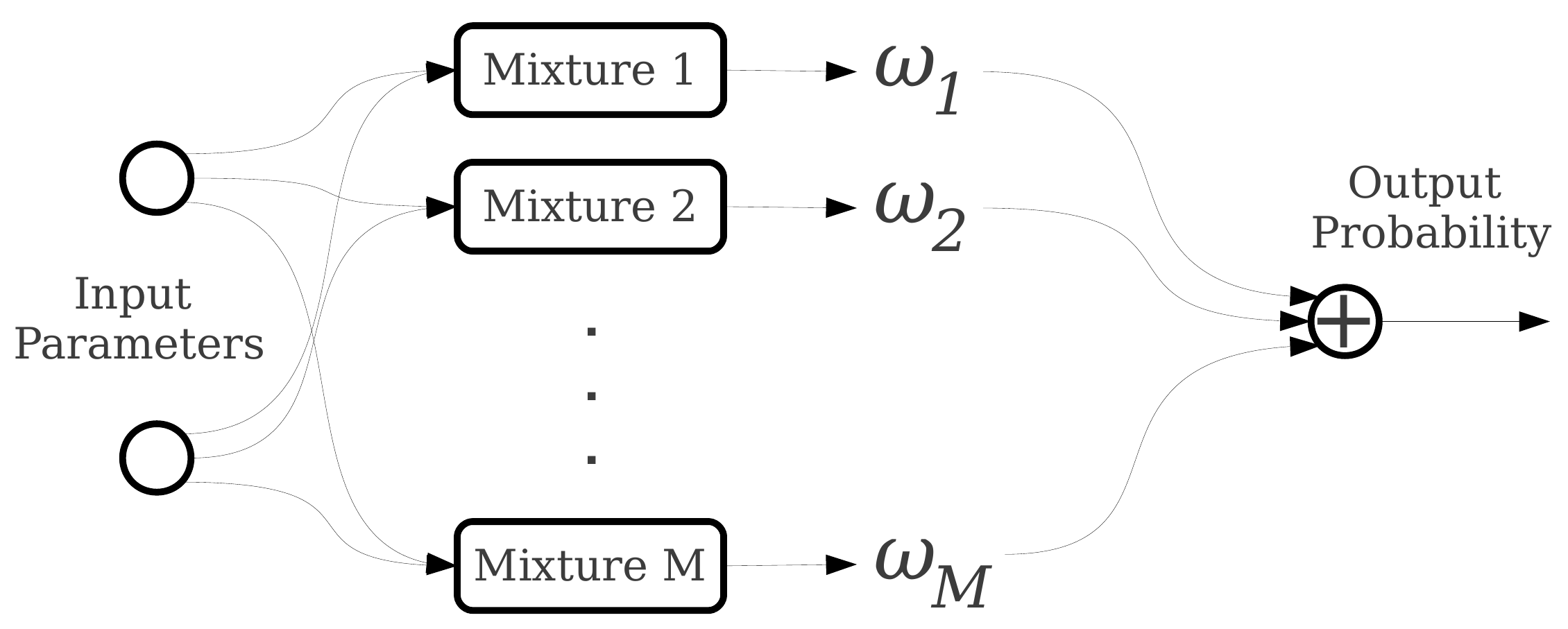}
\caption{Framework of the GMM}
\label{fig:gmmScheme}
\end{center}
\end{figure}

The Gaussian Mixture Model (GMM) is a classic parametric model used in many pattern recognition techniques to represent multivariate probability distribution. Any general distribution is approximated by sum of weighted Gaussian distributions. The overall framework is visualized in Figure~\ref{fig:gmmScheme}.

In the estimation process, we use GMM-based density function to calculate output probability of a feature $ x $ using a weighted combination of multi-variate Gaussian densities. Briefly, the GMM is a weighted sum of $ D $ component densities and given by the equation
\begin{equation}
\mathit{GMM}_{\lambda}(x) = \sum_{i=1}^{D}\omega_{i}N_{i}\left(x\right),
\end{equation}
where $ N_{i}(x) $ is the multi-variate Gaussian distribution and defined as 

\begin{equation}
N_{i}(x) = \frac{1}{\sqrt{2\pi^{D}|\sum|}}\,\exp\Big(-\frac{1}{2}(x-\mu_{i})^{T}{\sum}^{-1}(x-\mu_{i})\Big)
\end{equation}
and $ \omega_{i} $ is the mixture weight corresponding to the $ i $-th mixture and satisfies

\begin{equation}
\sum_{i=1}^{D}\omega_{i}=1\:\text{and}\:\omega_{i}\geq0.
\end{equation}
$ \lambda $ is the model and described by

\begin{equation}
\lambda=\left\{\omega_{i},\mu_{i},\sum\right\},
\end{equation}
where $ \mu_{i} $ is the mean of the $ i $-th Gaussian mixture and $ \sum $ is the diagonal covariance matrix. 

In the GMM context, a speaker's speech is characterized by $ D $ acoustic classes representing broad phones in language. The probabilistic modeling of an acoustic class is important since there is variability in features coming from the same class due to variations in pronunciation and articulation. Thus, the mean vector $ \mu_{i} $ represents the average features for the $ i $-th acoustic class and the covariance matrix $ \sum $ models the variability of features within the acoustic class. In this model, the covariance matrix is typically assumed to be diagonal because of computational concern. The GMM parameters are usually estimated by a standard iterative parameter estimation procedure, which is a special case of the \textit{Expectation-Maximization} (EM) algorithm and the initialization is provided by the \textit{Vector Quantization} (VQ) method.

\section{GMM-Based Probabilistic Mapping}
\label{subsec:GMM}

The Gaussian mixture model (GMM) estimator of \cite{Stylianou1998,Agiomyrgiannakis2007} is a soft mapping (SM) from observable source ${\cal X}$ to hidden source ${\cal Y}$ with an optimal linear transformation in the minimum mean square error (MMSE) sense. This mapping can be formulated as the MMSE estimator from the observable source to the hidden source,
\begin{equation}
\hat{\m{y}}_k^{s} = \sum_{l=1}^{L} p(\gamma_l|\m{x}_k) [{\mu}_{y,l} + \m{C}_{yx,l} (\m{C}_{xx,l})^{-1}(\m{x}_k-{\mu}_{x,l})],
\end{equation}
where $\gamma_l$ is the $l$-th Gaussian mixture and $L$ represents the total number of Gaussian mixtures. The vectors ${\mu}_{x,l}$ and ${\mu}_{y,l}$ are respectively the centroids for the $l$-th Gaussian for sources ${\cal X}$ and ${\cal Y}$, $\m{C}_{xx,l}$ is the covariance matrix of source ${\cal X}$ in the $l$-th Gaussian, and $\m{C}_{yx,l}$ is the cross-covariance matrix of sources ${\cal X}$ and ${\cal Y}$ for the $l$-th Gaussian mixture. The probability of the $l$-th Gaussian mixture given the observation $\m{x}_k$ is defined as the normalized Gaussian pdf as,
\begin{equation}
p(\gamma_l|\m{x}_k) = \frac{{\cal N}(\m{x}_k;{\mu}_{x,l}, \m{C}_{xx,l})}{\sum_{m=1}^L {\cal N}(\m{x}_k;{\mu}_{x,m}, \m{C}_{xx,m})}.
\end{equation}	

The GMM estimator can also be formulated as a hard mapping (HM) from the observable source ${\cal X}$ to the hidden source ${\cal Y}$ as,
\begin{equation}
\hat{\m{y}}_k^{h} = p(\gamma_{l^*}|\m{x}_k) [{\mu}_{y,l^*} + \m{C}_{yx,l^*} (\m{C}_{xx,l^*})^{-1}(\m{x}_k-{\mu}_{x,l^*})],
\end{equation}
where $\gamma_{l^*}$ represents the most likely mixture component, that is, 

\begin{equation}
l^* = \arg\max_l p(\gamma_l|\m{x}_k).
\end{equation}

\section{Enhancement Framework}
\label{subsec:EnhFrame}

Let us consider having two simultaneously recorded TM and AM speech, which are represented as $s_T[n]$ and $s_A[n]$, respectively. Source-filter decomposition through the linear prediction filter model of speech can be defined as,
\begin{eqnarray}
\label{eqn:synthesis}
S_{TT}(z) & = & \frac{1}{W_T(z)} R_T(z) \\
S_{AA}(z) & = & \frac{1}{W_A(z)} R_A(z),
\end{eqnarray}
where $W_T(z)$ and $W_A(z)$ are the inverse linear prediction filters, and $R_T(z)$ and $R_A(z)$ are the source excitation spectra for the TM and AM speech, respectively. Then we can define the TM speech enhancement problem as finding two mappings, the first one from TM spectra to AM spectra, and the second one from TM excitation to AM excitation,
\begin{eqnarray}
\widehat{W}_A(z) &=& \varphi_S (W_T(z) | \Lambda_S^{TA} ),\\
\widehat{R}_A(z) &=& \varphi_E (R_T(z) | \Lambda_E^{TA} ),
\end{eqnarray}
where $\Lambda_S^{TA}$ and $\Lambda_E^{TA}$ are general correlation models of TM and AM spectral envelopes and excitation. These joint correlation models can be extracted using a simultaneously recorded training database. Replacing the TM speech spectra and excitation with the estimates,
\begin{equation}
\label{eqn:lsfMapping}
\widehat{S}_{AA}(z) = \frac{1}{\widehat{W}_A(z)} \widehat{R}_A(z),
\end{equation}
is expected to enhance the perceived quality of the TM speech. Similarly, we can replace only TM spectra or excitation and also none of them as~\eqref{eqn:synthesis} to see detailed the effect of each mapping individually. 
\begin{eqnarray}
\label{eqn:otherSyntesis}
\widehat{S}_{AT}(z) = \frac{1}{\widehat{W}_A(z)} {R}_T(z) \\
\widehat{S}_{TA}(z) = \frac{1}{{W}_T(z)} \widehat{R}_A(z)
\end{eqnarray}

\section{Spectral Envelope Enhancement}
\label{subsec:EnvMapping}

In this study, the line spectral frequency (LSF) feature vector representation of the linear prediction filter is used to model spectral envelope. The TM and AM spectral representations are extracted as $16$th order linear prediction filters over $10$~ms time frames. We define the elements of this representation at time frame $k$ as column vectors $\m{x}^s_k$ and $\m{y}^s_k$, respectively representing the TM spectral envelope as an observable source ${\cal{X}}^s$ and AM spectral envelope as a hidden source ${\cal{Y}}^s$. Throat-microphone recordings reveal certain speech attributes, and deliver varying perceptual quality for different sound vocalizations,
such as nasals, stops, fricatives. Hence an acoustic phone dependent relationship between throat- and acoustic-microphone speech can be formulated to value the attributes of the throat-microphone speech. In order to explore such a relationship between throat- and acoustic-microphone speech, we first define a phone-dependent soft mapping (PDSM),
\begin{equation}
\label{eqn:PDSM}
 \hat{\m{y}}_k^{s|c} = \frac{1}{N} \sum_{n=1}^{N} \sum_{l=1}^{L_n} p(\gamma_{l}^{c_n} |\m{x}_k^h) [{\mu}_{y,l}^{c_n} + \m{C}_{yx,l}^{s|c_n} (\m{C}_{xx,l}^{s|c_n})^{-1} (\m{x}_k^s-{\mu}_{x,l}^{s|c_n})],
\end{equation}
where $N$ is the number of context elements and each phone $c_n$ has a separate GMM, which is defined by phone-dependent mean vectors and covariance matrices. The phones are set as the context $c_n$. 

Furthermore, a phone-dependent hard mapping (PDHM) can be defined as,
\begin{equation}
\label{eqn:PDHM}
 \hat{\m{y}}_k^{h|c} = \sum_{l=1}^{L^*} p(\gamma_{l}^{c^*} | \m{x}^h_k) [{\mu}_{y,l}^{h|c^*} + \m{C}_{yx,l}^{h|c^*} (\m{C}_{xx,l}^{h|c^*})^{-1} (\m{x}^h_k-{\mu}_{x,l}^{h|c^*})],
\end{equation}
where $c^*$ is the given phone, and $L^*$ is the total number of Gaussian mixtures for the phone class $c^*$. In this study we consider three different sources for the given context. The true context, $c^{T}$, is defined as the true phonetic class of the phone, which is considered as the most informative upper bound for the phone-dependent model. The likely context from the GMM, $c^{G}$, is defined as the most likely phonetic class, which can be extracted as,
\begin{equation}
 c^G = \arg\max_{c_n} {\cal N}(\m{x}_k;{\mu}_{x,l}^{c_n}, \m{C}_{xx,l}^{c_n}).
\end{equation}
Finally, the likely context from an HMM-based phoneme recognizer, $c^M$, is defined as the most likely phonetic class, which is decoded by an HMM-based phoneme recognition over the observable source ${\cal X}$.

\section{Excitation Enhancement}
\label{subsec:ExcMapping}

In the source-filter decomposition framework we observed that the TM and AM recordings exhibit significant differences at the excitation signal spectra, which appears to be an important source of the degradation in the TM voice quality. Hence, we model spectral envelope difference of the excitation signals as a spectral tilt vector, and we propose a new phone-dependent GMM-based spectral tilt mapping scheme to enhance TM excitation.

Let us first define a triangular filter-bank, which will help us to compute the average spectrum around a sequence of center frequencies,
\begin{equation}
w_b(n)=\left\{
\begin{array}{ll}
0 & n < f_{b-1}\;\;{\rm or}\;\; n > f_{b+1}\\
\frac{n-f_{b-1}}{f_b - f_{b-1}} & f_{b-1} \leq n \leq f_{b}\\
\frac{f_{b+1}-n}{f_{b+1} - f_b} & f_{b} < n \leq f_{b+1}
\end{array}\right.
\end{equation}
where $f_b$ is the $b$-th center frequency index and $w_b$ is the $b$-th triangular filter. We take number of bands as $B=8$ and let $b=0,1,\dots,B+1$, where $f_0=0$ and $f_{B+1}=N$ are taken as boundary frequency indexes. Then the average spectrum energy of the AM excitation signal is computed for frequency band $b$ as,
\begin{equation}
E_A(b) = \log \{ \sum_{n=1}^{N-1} w_b(n) |R_{A}(n)|^2 \}\;\; {\rm for}\;\; b=1,\dots,B,
\end{equation}
where $R_A$ is the $2N$-point DFT of the excitation signal, and $B$ is the total number of frequency bands. Similarly the average spectrum energies of the TM excitation can be computed and represented as $E_T$.

We can now define the spectral tilt vector between the TM and AM excitation signals as,
\begin{equation}
D(b) = E_{A}(b) - E_{T}(b) \;\; {\rm for}\;\; b=1,\dots,B.
\end{equation}
The spectral tilt vector is considered as the representation of the hidden source ${\cal{Y}}^e$ for the probabilistic excitation mapping. We define the spectral tilt vector at time frame $k$ as column vector $\m{y}^e_k$ representing the hidden source ${\cal{Y}}^e$. The observable source of the spectral envelope mapping, which is the $16$th order LSF feature vector $\m{x}^s_k$ of TM speech can be considered as a valuable observation also for the probabilistic excitation mapping. However we also consider excitation spectrum of the TM speech to be valuable. Hence we compute a cepstral feature vector representing the TM excitation spectrum as,
\begin{equation}
c_T(n) = \sum_{b=1}^B E_T(b) \cos(\pi n (b-1/2)/B),
\end{equation}
for $n=1,2,\dots,B-1$. We form the observable source vector of the excitation mapping as $\m{x}^e_k = [ \m{x}'^s_k \; c'_T]'$, where $c_T$ representing the cepstral column feature vector at frame $k$. Then, a phone-dependent mapping for the excitation enhancement of TM recordings is defined similarly as in (\ref{eqn:PDHM}) to estimate $\hat{\m{y}}_k^{e|c}$, or equivalently the spectral tilt vector $\widehat{D}$. Finally, the enhanced excitation spectrum can be estimated by tilting the TM excitation spectrum as following,
\begin{equation}
\widehat{R}_A^{\widehat{D}}(n) = \sum_{b=0}^{B+1} w_b(n) 10^{\widehat{D}_b} R_{T}(n) \;\;  n=1,\dots,N-1,
\end{equation}
where boundary spectral tilt values are taken as $\widehat{D}_0 = \widehat{D}_1$ and $\widehat{D}_{B+1} = \widehat{D}_B$.

Note that in processing of the excitation signals, a 2048-point DFT is used over 20~ms hamming windowed excitation signals with a frame shift of 10~ms. The enhanced excitation signal is reconstructed from the $\widehat{R}_A^{\widehat{D}}$ spectrum with inverse DFT and overlap-and-add schemes.
\chapter{Experimental Evaluations}
\label{ch:ch3_experiments}

We perform experiments on a synchronous TM and AM database which consists of two male speakers namely, M1 and M2. The latter one is recorded with a new IASUS-GP3 headset. Also, the AM data comes from Sony electret condenser tie-pin microphone. Each speaker utters 799 sentences that are recorded simultaneously at 16-kHz sampling rate. At the training stage, codebooks are established via varying number of Gaussian mixtures model using one-fold cross validation. In other words, we use 720 sentences as training data and the rest of the recordings as test data in our speaker dependent mapping schemes.

Experimental evaluations are divided into two sub-groups. As it discussed in Chapter~\ref{ch:ch2_enhancement} that speech can be separated into two independent parts, namely source and filter. Thus, firstly, different mapping schemes for the enhancement of filter are applied, then, excitation improvement is carried out with a branch of experiments to emphasize  their individual effect more detailed. For envelope enhancement, we use a database from one male speaker only namely M1 and for excitation enhancements, both male records (M1 and M2) are used for comprehensive analysis. 

\section{Observations on Throat-Microphone Speech Attributes}
\label{sec:obser}

It can be helpful to analyze the phones according to how they are articulated in oral cavity. The articulation of different phones come with its distinct character in terms of resonance shaping. Although they differ in realization across individual speakers, the tongue shape and positioning in the oral cavity do not change significantly. Since, the throat-microphone captures a reliable low-frequency energy, it can represent baseband spectrum, such as nasals and voice bars, sufficiently well. In articulatory phonetics, manner of articulation is very important parameter for classification of phones. It describes the degree of narrowing in the oral cavity and certain acoustic or perceptual characteristics. For example, the phone /\textit{n}/ and /\textit{k}/ have same manner of articulation because they are articulated by the rapid release of a complete oral closure. Likewise, /\textit{s}/ and /\textit{z}/ have same manner of articulation and are articulated by forming a constriction that causes a turbulence in the flowing air so they produce a hissing sound. 

In Table~\ref{tbl:attribute} we collect the average LSD scores between the acoustic and throat spectral envelopes, respectively $W_A(\omega)$ and $W_T(\omega)$, for the main phonetic attributes. The two minimum LSD scores occur for the nasals and stops, and the fricatives yield the maximum LSD score. Note that, nasals realized over closure of nasal cavity such as {\em/m/} have smallest distortion, and fricatives realized over the friction of narrow-stream turbulent air such as {\em/s/} have largest distortion due to its high-frequency energy. Clearly, the mapping of the throat-microphone speech spectra to the acoustic-microphone speech spectra is harder for the fricatives than for the nasals. That is one of the main reasons that we investigate a context-dependent mapping for the enhancement of throat-microphone speech.

\begin{center}
\captionof{table}{The average LSD scores of the data M1 between throat- and acoustic-microphone spectrums for different phonetic attributes with relative occurrence frequencies in the test database.}
\label{tbl:attribute}
\begin{tabular}{c|cc@{}}
\toprule[1.5pt]
\multicolumn{1}{c}{\head{Attribute}} & \multicolumn{1}{c}{\head{Freq}} & \head{LSD (dB)}\\
\midrule
Nasals & 9.27 & 5.58 \\
Stops & 16.94 & 6.27 \\
Liquids & 9.59 & 7.05 \\
Back Vowels & 16.18 & 7.22 \\
Front Vowels & 13.93 & 7.65 \\
Glide & 2.36 & 7.81 \\
Affiricate & 2.72 & 9.54 \\ 
Fricatives & 11.10 & 11.81 \\
\bottomrule[1.5pt]
\end{tabular}
\end{center}

\section{Objective Evaluations}
\label{subsec:ObjEval}

Evaluations of the TM speech enhancement are performed with two distinct objective metrics, the logarithmic spectral distortion (LSD) and the perceptual evaluation of wide-band speech quality (PESQ) metrics. The logarithmic spectral distortion (LSD) is a widely used metric for spectral envelope quality assessment. The LSD metric assesses the quality of the estimated spectral envelope with respect to the original wide-band counterpart, and is defined as

\begin{equation}
d_{LSD} =\sqrt{{\frac{1}{2\pi}\int_{-\pi}^{\pi} \left (10 \log \frac{|W_A(\omega)|^2}{|\hat{W}_A(\omega)|^2} \right )^2 d\omega}}
\end{equation}
where $W_A(\omega)$ and $\hat{W}_A(\omega)$ represent the original and estimated acoustic spectral envelopes, respectively. The ITU-T Standard PESQ \cite{wb-pesq2005} is employed as the second objective metric to evaluate the perceptual quality of the enhanced throat-microphone speech signal, which is constructed using the estimated spectral envelope and the excitation signal of the throat-microphone speech. The PESQ algorithm predicts subjective opinion scores of a degraded speech sample from 4.5 to −0.5 (higher score indicates better quality).

In phonetics and linguistics, a phone is defined as a unit of speech sound that is the smallest identifiable part found in a stream of speech. Since it is pronounced in a defined way, we can regard it as a context data. From the Table~\ref{tbl:metubet}, there are 37 different phones despite of the fact that 29 letters are available in Turkish language. The recordings are phonetically transcribed using the Turkish phonetic dictionary METUbet and the phone level alignment is performed using forced-alignment and visual inspection. In \cite{Salor2002}, it is developed a new letter-to-phone conversion rule set that is based on the phonetic symbol set of Turkish language. These rules are formed by observing the phonetic transcriptions of the letters in the dictionary and determining the phonetic conditions in which they appear \cite{Salor2002}. The choice of symbol formatting in METUbet is similar to that used within ARPAbet for American English. The METUbet phonetic alphabet is given in Table~\ref{tbl:metubet} where phones are categorized into 8 different manner of articulation.

\begin{table}[H]
\centering
\caption{The Turkish METUbet phonetic alphabet with classification into 8 articulation attributes.}
\label{tbl:metubet}
\begin{center}
\as{1.1}
\begin{tabular}{||l|l||l|l||l|l||} 
\hline 
\multicolumn{2}{||c||}{\textbf{Back Vowels}}	& \multicolumn{2}{c||}{\textbf{Stops}}	& \multicolumn{2}{c||}{\textbf{Fricatives}} \\ \hline
AA &	{\bf a}n{\i} 			& B  &  {\bf b}al			& H &	{\bf h}asta \\
A  &	l{\bf a}f 				& D  & {\bf d}e{\bf d}e		& J &	m\"{u}{\bf j}de \\
I  &	{\bf{\i}}s{\bf{\i}}			& GG & kar{\bf g}a		& F &	{\bf f}as{\i}l \\
O  &	s{\bf o}ru				& G &	{\bf g}en\c{c}		& S &	{\bf s}e{\bf s} \\
U  &	k{\bf u}lak				& KK & a{\bf k}{\i}l 		& SH & a{\bf \c{s}}{\i} \\ \cline{1-2}\cline{1-2}
\multicolumn{2}{||c||}{\textbf{Front Vowels}}& K &	{\bf k}edi			& VV & {\bf v}ar \\ \cline{1-2}
E  &	{\bf e}lma				& P &	i{\bf p}			& V &	ta{\bf v}uk \\
EE &	der{\bf e}				& T &	\"{u}{\bf t}\"{u}		& Z &	a{\bf z}{\i}k \\ \cline{3-4} \cline{3-4}
IY &	s{\bf i}m{\bf i}t			&\multicolumn{2}{c||}{\textbf{Liquids}}& ZH &	yo{\bf z} \\ \cline{5-6}\cline{5-6} \cline{3-4}
OE &	{\bf \"{o}}rt\"{u}			& LL &	ku{\bf l}		& \multicolumn{2}{c||}{\textbf{Affiricates}} \\ \cline{5-6}
UE &	{\bf \"{u}}mit			& L &	{\bf l}ey{\bf l}ek		& C &	{\bf c}am \\ \cline{1-2}\cline{1-2}
\multicolumn{2}{||c||}{\textbf{Nasals}}	& RR &	I{\bf r}mak		& CH & se{\bf \c{c}}im \\ \cline{5-6}\cline{5-6}\cline{1-2}
M &	da{\bf m} 				& RH &	bi{\bf r}		& \multicolumn{2}{c||}{\textbf{Glide}} \\ \cline{5-6}
NN &	a{\bf n}{\.i}			& R &	{\bf r}af			& Y &	{\bf y}at \\
N &	s\"{u}{\bf n}g\"{u}		&  & & & \\ \hline

\end{tabular}
\end{center}
\end{table}

As Salor et al. suggest in \cite{Salor2002}, the Turkish phone GH, soft g, has not been used for transcription and recognition, since it is used for lengthening of the previous vowel sound. 

In machine learning, taking the contextual information into account undoubtedly improves the performance of any system. Since articulation of speech can be approximated as a stationary process, knowing the piece of a speech that surrounds a particular word can contribute its full analysis. Therefore, the proposed enhancement schemes take phone information into consideration as probabilistic mappings.

\subsection{Envelope Enhancements}
\label{ssec:envelopeEnh}

Table~\ref{tbl:ObjRes1} presents the average LSD scores between the estimated filter $\hat{W}_A(z)$ and the original acoustic filter ${W}_A(z)$, and the average PESQ scores between the enhanced throat-microphone recordings and the original acoustic-microphone recordings. Note that for increasing the PESQ, the LSD scores decrease in a consistent manner.

\begin{center}
\captionof{table}{The average LSD and PESQ scores for different mapping schemes for enhancement of the throat-microphone recordings.}
\label{tbl:ObjRes1}
\begin{tabular}{r|cc@{}}
\toprule[1.5pt]
\multicolumn{1}{c}{} & \head{LSD} & \head{PESQ}  \\
\multicolumn{1}{c}{} & \head{(dB)} & \head{(MOS-LQO)} \\
\midrule
\head{PDHM-G} & 3.92 & 1.27 \\ 
\head{HM} & 3.80 & 1.29  \\ 
\head{SM} & 3.66 & 1.34 \\ 
\head{PDSM} & 3.65 & 1.36 \\
\head{PDHM-M} & 3.48 & 1.38 \\ 
\head{PDHM-T} & 3.18 & 1.43 \\
\bottomrule[1.5pt]
\end{tabular}
\end{center}

$ {} $ \\
The number of mixture components for the phone-dependent hard and soft mapping schemes are set as $L_n=16$ for all phones. Similarly the number of mixture components for the GMM based hard and soft mapping schemes is set as $L=256$. The worst performing scheme is observed as the phone-dependent hard mapping when phone recognition is performed with the GMM classifier (PDHM-G). The hard and soft mapping schemes, respectively HM and SM, achieve some performance improvement over the PDHM-G mapping. The best LSD and PESQ scores are attained with the phone-dependent hard mapping when the true phone class is known (PDHM-T). Also, the phone-dependent soft mapping (PDHM) performs close to the soft mapping (SM) scheme. 

The phone-dependent hard mapping with the HMM-based phone recognition (PDHM-M) attains a performance improvement and performs closest to the PDHM-T mapping scheme. The HMM-based phone recognition for the PDHM-M mapping is performed with 3-state and 256-mixture density phone level HMM recognizer, which is trained over the throat-microphone recordings of the 11 male speakers of the TAM database in \cite{Erzin2009}. Note that the test recordings in this study have been excluded from the training set of the phone level HMM recognizer. The HMM-based phone recognizer attains 62.22\% correct phone recognition over the test database. We observe significant performance improvement when the true phone class is known to the phone-dependent hard mapping (PDHM-T) scheme. Furthermore the phone-dependent hard mapping with a reliable phone recognition, in this case the PDHM-M mapping, attains the best blind estimation for the spectral envelope to enhance the throat-microphone recordings.

\begin{center}
\captionof{table}{The average PESQ scores for different mapping schemes using acoustic residual.}
\label{tbl:ObjRes2}
\begin{tabular}{r|c@{}}
\toprule[1.5pt]
\multicolumn{1}{c}{} & \head{PESQ} \\
\multicolumn{1}{c}{} & \head{(MOS-LQO)} \\ 
\midrule
\head{PDHM-G}  & 1.66 \\ 
\head{HM}  & 1.75 \\ 
\head{SM} & 1.97 \\ 
\head{PDSM}  & 2.02 \\ 
\head{PDHM-M}  & 2.16 \\ 
\head{PDHM-T} & 2.53 \\ 
\bottomrule[1.5pt]
\end{tabular}
\end{center}

$ {} $ \\ 
The throat-microphone recordings have a lower bandwidth at low-frequency bands compared to the reference acoustic-microphone recordings. Since the perceived intelligibility is poor for the throat-microphone recordings, the average PESQ scores stay at low values. In order to isolate the degradation, which is introduced by the throat source $R_T(z)$, we consider the case with the acoustic source $R_A(z)$ and throat filter ${W}_T(z)$ as a degraded speech signal. In this case we synthesized an enhanced speech signal using the estimated filter $\hat{W}_A(z)$ and the acoustic source $R_A(z)$. Table~\ref{tbl:ObjRes2} presents the average PESQ scores for this investigation. Note that the PESQ results are higher compared to Table~\ref{tbl:ObjRes1}. Furthermore the phone-dependent hard mapping PDHM-M scheme has the highest PESQ improvement.

\subsection{Excitation Enhancements}
\label{ssec:excitationEnh}

In this part, we consider only phone-dependent mapping schemes with two possible sources of the phone. First source is obtained by forced-alignment, and the second one is picked up from an HMM-based phone recognition system over the observable TM source. The phone recognition performances for the M1 and M2 speakers are obtained respectively as 62.22\% and 61.07\%.

We first investigate possible best case scenarios for the enhancement of TM recordings when AM speech data are available. Table~\ref{tbl:ObjRef} presents average PESQ scores for the four scenarios, where reference condition is always the AM recordings and a form of the TM recordings are synthesized with the given excitation and filter models. The first row presents the average PESQ scores between TM and AM recordings for both speakers M1 and M2. The second row presents the average PESQ scores of the source-filter synthesis when the TM filter is replaced by the AM filter. We observe similar PESQ score improvements for both speakers, and these can be considered as best case improvements for the enhancement of the spectral envelope. The last two rows of Table~\ref{tbl:ObjRef} presents average PESQ scores when AM filter is used and TM excitation is tilted with the original spectral tilt vector with linearly spaced frequency bands ($D_{lin}$) and with mel-scaled frequency bands ($D_{mel}$). Note that when the TM excitation is tilted with the original spectral tilt vector we observe high PESQ score improvements. However, we do not discover any improvement with the use of mel-scaled frequency bands in the computation of the spectral tilt vectors. Hence, we keep using the linearly spaced frequency bands in the remaining parts of the experimental evaluations.

\begin{center}
\captionof{table}{The average PESQ scores for evaluation of the targeted excitation and filter enhancement strategies.}
\label{tbl:ObjRef}
\begin{tabular}{cc|cc@{}}
\toprule[1.5pt]
\multicolumn{2}{c}{} & \multicolumn{2}{c}{\head{PESQ}} \\
\multicolumn{2}{c}{} & \multicolumn{2}{c}{\head{(MOS-LQO)}} \\
\cmidrule{3-4}
\head{Exc.} & \multicolumn{1}{c}{\head{Filter}} & \multicolumn{1}{c}{\head{M1}} & \head{M2} \\
\midrule
$R_T$ & $W_T$ & 1.22 & 1.42 \\
$R_T$ & $W_A$ & 1.70  & 1.74 \\
$\widehat{R}_A^{D_{lin}}$ & $W_A$ & 2.23 & 2.72 \\
$\widehat{R}_A^{D_{mel}}$ & $W_A$ & 2.26 & 2.63 \\
\bottomrule[1.5pt]
\end{tabular}
\end{center}

\begin{figure}[!ht]
\centering
\includegraphics[scale=0.6]{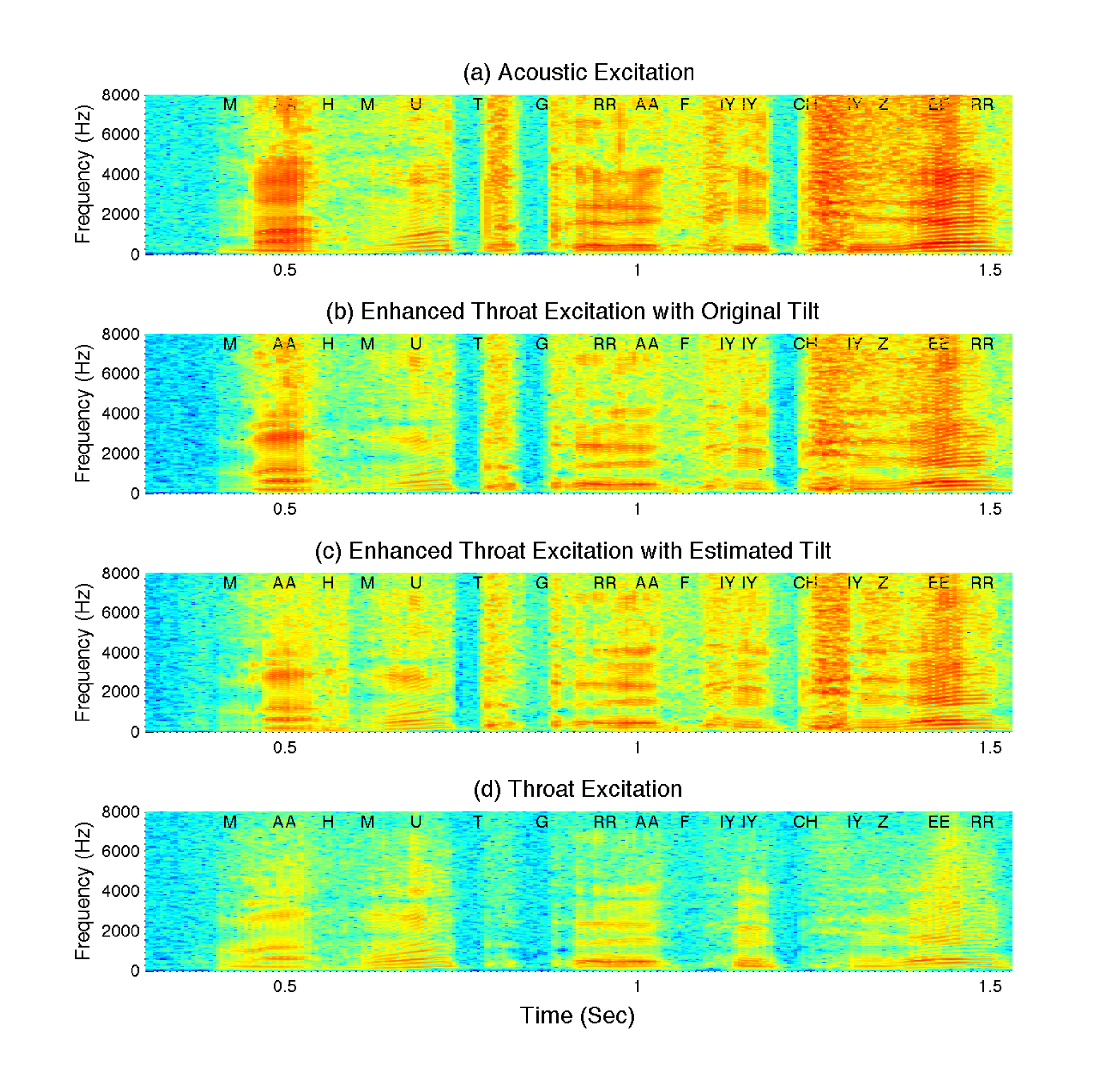}
\caption{Sample spectrograms of (a) AM excitation, (b) enhanced TM excitation with the original spectral tilt, (c) enhanced TM excitation with the estimated spectral tilt, and (d) TM excitation.
\label{fig:sx450}}
\end{figure}

$ {} $ \\
Figure~\ref{fig:sx450} presents spectrograms of the excitation signals to emphasize the effect of spectral tilt. Clearly, the third from the top (c) is the TM excitation, which is enhanced by the estimated spectral tilt vector using the proposed probabilistic mapping. Note that enhancement of the TM excitation significantly compensates spectral energy distribution with respect to the AM excitation.

Table~\ref{tbl:ObjRes} presents PESQ scores for different excitation and spectral envelope mapping schemes that we use for the enhancement of TM recordings. First two columns define excitation and filter mapping schemes. The PESQ scores are presented with respect to the source of phone information, either force alignment or HMM-based phone recognition, and speaker information M1 and M2. The first observation is on the source of the phone information. Given the PESQ scores with the reliable force aligned phone information, the PESQ scores with the HMM-based phone recognition information do not degrade significantly. For example, the first row of the results, where only the spectral envelope is mapped, PESQ score drop is from 1.59 to 1.58 for speaker M2 with the use of phone recognition information. The second observation is on the attained PESQ score improvements with the proposed excitation and spectral envelope mapping schemes. The PESQ scores 1.22  and 1.42 of the TM recordings as reported in Table~\ref{tbl:ObjRef} respectively for speakers M1 and M2 are improved to 1.46 and 1.58 with the spectral envelope mapping, and to 1.61 and 1.86 with the excitation and spectral envelope mappings when phone information is extracted from the HMM-based phone recognizer. Here surprising observation is the contribution of the excitation enhancement, which brings a higher improvement than the spectral envelope enhancement. Finally the third observation is on the sole contribution of the excitation mapping, where last row of Table~\ref{tbl:ObjRes} presents PESQ scores with the proposed excitation mapping when TM filter is used. We do not observe PESQ score improvements for the sole use of excitation mapping, hence we can say that the proposed excitation mapping is significant when used with the spectral envelope mapping.

\begin{center}
\captionof{table}{The average PESQ scores for different excitation and spectral envelope mapping schemes. 
\label{tbl:ObjRes}}
\begin{tabular}{C{1.2cm}C{1.2cm}cC{1.2cm}C{1.2cm}cC{1.2cm}C{1.2cm}c}
\toprule[1.5pt] 
{} & {} & {} & \multicolumn{5}{c}{\head{PESQ (MOS-LQO)}} & {} \\
\cmidrule{4-8}
{} & {} & {} & \multicolumn{2}{c}{\head{Force-Align.}} & {} & \multicolumn{2}{c}{\head{Phone Recog.}} & {} \\
\cmidrule{4-8}
\head{Exc.} & \head{Filter} & {} & \head{M1} & \head{M2} & {} & \head{M1} & \head{M2} & {} \\ 
\midrule
$R_T$ & $\widehat{W}_A$ & \vline & 1.53 & 1.59 & \vline & 1.46 & 1.58 & \vline \\
$\widehat{R}_A^{\widehat{D}}$ & $\widehat{W}_A$ & \vline & 1.65 & 1.90 & \vline & 1.61 & 1.86 & \vline \\
$\widehat{R}_A^{\widehat{D}}$ & $W_T$ & \vline & 1.34 & 1.43 & \vline & 1.28 & 1.40 & \vline \\  
\bottomrule[1.5pt]
\end{tabular}
\end{center}

$ {} $ \\
From the objective evaluations of both envelope and excitation enhancements, we can clearly notice that phone-dependent mapping is the best enhancement model in terms of intelligibility and spectral distortion. This is because of the accuracy in context information that suits the nature of speech enhancement well. In other words, if we know the precede and past of a particular speech very well, we can built more qualified model for the TM enhancement. 

\section{Subjective Evaluations}
\label{subsec:SubjEval}

Since the reported PESQ scores stay at low values, a subjective evaluation of the proposed throat-microphone speech enhancement techniques is necessary to check whether the objective score improvements are subjectively perceivable. We performed a subjective A/B comparison test to evaluate the proposed enhancement techniques. During the test, the subjects are asked to indicate their preference for each given A/B test pair of sentences on a scale of (-2; -1; 0; 1; 2), where the scale corresponds to {\em strongly prefer A}, {\em prefer A}, {\em no preference}, {\em prefer B}, and {\em strongly prefer B}, respectively. The subjective tests are divided into two parts likewise in objective evaluations.

\subsection{Envelope Enhancements}
\label{subsec:EnvMaps}

The subjective A/B test of envelope mapping schemes includes 21 listeners, who compared 20 sentence pairs randomly chosen from our test database to evaluate 5 conditions. The acoustic-microphone speech condition is compared to all conditions with 1 pair. The throat-microphone speech condition is compared to all three enhancement schemes with 2 pairs. The GMM-based soft mapping scheme is compared to the phone-dependent hard mapping schemes PDHM-T and PDHM-M with 3 pairs. Finally, the PDHM-T scheme is compared to the PDHM-M scheme with 3 pairs.

Table~\ref{tbl:SubRes1} presents the average subjective preference results. The rows and the columns of these tables correspond to A and B conditions of the A/B pairs, respectively. Also, the average preference scores that tend to favor B are given in bold to ease visual inspection. Speech samples from the subjective A/B comparison test are available for online demonstration \cite{Turan2012Web}.

\begin{center}
\captionof{table}{The average preference results of the subjective A/B
pair comparison test for envelope mapping}
\label{tbl:SubRes1}
\begin{tabular}{@{}C{1.2cm}C{2.2cm}C{1.2cm}C{1.2cm}C{1.2cm}C{1.2cm}}
\toprule[1.5pt]
 & & \multicolumn{4}{c}{\head{B}} \\
\cmidrule{3-6}
 & \head{A} & {\bf 1} & {\bf 2} & {\bf 3} & {\bf 4}  \\
\cmidrule{2-6}
{\bf 1} & Acoustic & {\bf 0.05} 	&  &  &   \\
{\bf 2} & Throat & {\bf 1.93}  	& &  &   \\
{\bf 3} & SM & {\bf 1.93}  		& -0.57 	&  &  \\
{\bf 4} & PDHM-T & {\bf 1.83} 	& -1.12	& -0.49 	&  \\
{\bf 5} & PDHM-M & {\bf 1.83}	& -0.83	& -0.27	& {\bf 0.54}\\
\bottomrule[1.5pt]
\end{tabular}
\end{center}

$ {} $ \\
All the three enhancement schemes yield a perceivable difference compared to the throat-microphone speech. Among the three enhancement schemes, the PDHM-T, which uses the true phone class, has the highest perceivable improvement. The proposed PDHM-M scheme has the second best apprehensible improvement, which is inline with the objective evaluations.

\subsection{Excitation Enhancements}
\label{subsec:ExcMaps}

The subjective A/B test includes 33 listeners, who compared 29 sentence pairs randomly chosen from our test database of speaker M2 to evaluate 6 conditions. The AM and TM speech conditions are compared to all conditions with 1 pair. The second row of the Table~\ref{tbl:ObjRef} ($R_T$, $W_A$) defines a target condition for the best possible spectral envelope mapping, which is evaluated as the third condition. 

\begin{center}
\captionof{table}{The average preference results of the subjective A/B
pair comparison test for excitation mapping}
\label{tbl:SubRes2}
\begin{tabular}{@{}C{1.2cm}C{2.2cm}C{1.2cm}C{1.2cm}C{1.2cm}C{1.2cm}C{1.2cm}}
\toprule[1.5pt]
\multicolumn{2}{c}{~} & \multicolumn{5}{c}{\head{B}} \\
\cmidrule{3-7}
\multicolumn{1}{c}{~} & \head{A} & {\bf 1} & {\bf 2} & {\bf 3} & {\bf  4} & {\bf 5}  \\
\cmidrule{2-7}
{\bf 1} & Throat 				& 0.00  &  &  &  & \\
{\bf 2} & Acoustic 				& -1.97 & {\bf 0.06} &  &  &  \\
{\bf 3} & $R_T$, $W_A$ 			& -1.21 & {\bf 1.70} &  &  &  \\
{\bf 4} & $\widehat{R}_A^{D}$, $W_A$ 	& -1.85 & {\bf 1.00} & -1.40 &  & \\
{\bf 5} & $R_T$, $\widehat{W}_A$ 		& -0.79 & {\bf 1.82} & {\bf 0.34}  & {\bf 1.72} &  \\ 
{\bf 6} & $\widehat{R}_A^{\widehat{D}}$, $\widehat{W}_A$ & -1.41 & {\bf 1.49} & -0.57 & {\bf 0.69} & -1.12 \\
\bottomrule[1.5pt]
\end{tabular}
\end{center}

$ {} $ \\ 
The fourth condition is defined as the target for the best possible excitation and spectral envelope mapping ($\widehat{R}_A^{D}$, $W_A$), which is the third row of the Table~\ref{tbl:ObjRef}. The fifth and sixth conditions are set as the two proposed enhancement schemes, the sole spectral envelope mapping ($R_T$, $\widehat{W}_A$) and the excitation and spectral envelope mappings ($\widehat{R}_A^{\widehat{D}}$, $\widehat{W}_A$), which are respectively the first two rows of the Table~\ref{tbl:ObjRes}.  Conditions 3-6 are compared to each other with 3 pairs.

Table~\ref{tbl:SubRes2} presents the average subjective preference results. The rows and the columns of Table~\ref{tbl:SubRes2} correspond to A and B conditions of the A/B pairs, respectively. Also, the average preference scores that tend to favor B are given in bold to ease visual inspection. Speech samples from the subjective A/B comparison test are available for online demonstration \cite{Turan2013Web}.

The proposed excitation and spectral envelope mapping scheme ($\widehat{R}_A^{\widehat{D}}$, $\widehat{W}_A$), which is the condition 6, has perceivable improvements compared to all conditions except the AM recordings and the best target mapping ($\widehat{R}_A^{D}$, $W_A$). Furthermore it is significantly preferred over the sole spectral envelope mapping ($R_T$, $\widehat{W}_A$) with a preference score 1.82, and it is the second closest condition to the AM recordings after the best target mapping with a preference score 1.49. 
\chapter{Conclusion}
\label{ch:conc}

In this thesis, we introduce a new phone-dependent GMM-based spectral envelope mapping scheme to enhance throat-microphone speech using joint analysis of throat- and acoustic-microphone recordings. The proposed spectral mapping scheme performs the minimum mean square error (MMSE) estimation of the acoustic-microphone spectral envelope within the phone class neighborhoods. Objective and subjective experimental evaluations indicate that the phone-dependent spectral mapping yields perceivable improvements over the state-of-the-art context independent mapping schemes. Overall, the proposed phone-dependent spectral mapping, PDHM-M, introduces a significant intelligibility improvement over the throat-microphone speech. However, there is still a big room to further improve the perceive quality by modeling the source excitation signal of the throat-microphone recordings. 

In the source-filter model of vocal tract, we observed that the spectral envelope difference of the excitation signals of TM and AM speech is an important source of the degradation in the throat-microphone voice quality. Thus, we model spectral envelope difference of the excitation signals as a spectral tilt vector using the same joint structure of TM and AM. Again, objective and subjective experimental evaluations indicate that the correction of the TM excitation spectrum has a strong potential to improve intelligibility for the TM speech. Although the proposed excitation mapping achieves a significant improvement (1.86 PESQ score for speaker M2) within the strong potential of correcting TM excitation with a spectral tilt (2.72 PESQ score of the best target mapping for speaker M2), there is still some unattempted improvement. We have to consider that incorporating temporal dynamics of the spectral tilt to the probabilistic mapping may attain further improvements for TM speech enhancement.

\bibliographystyle{IEEEtran}

\end{document}